# SIMULATION OF COUNTER FLOW PEDESTRIAN DYNAMICS IN HALLWAYS USING SPHEROPOLYGONS

**Fernando Alonso-Marroquin[1], Celia Lozano[2], Álvaro Ramírez-Gómez[3], and Jonathan Busch[1]**

[1]*School of Civil Engineering, The University of Sydney, Sydney, NSW, Australia*
[2] *School of Applied Mathematics and Physics, The University of Navarra, Spain*
[3] *Universidad Politécnica de Madrid, Spain*
[*] *Corresponding author – Fernando Alonso-Marroquin: fernando.alonso@sydney.edu.au*

We developed a method for simulating pedestrian dynamics in a large, dense crowd. Our numerical model calculates pedestrian motion from Newton's second laws, taking into account visco-elastic contact forces, contact friction, and ground reaction forces. In our computer simulation, non-spherical shapes (spheropolygons) modelled the positions of the chest and arms in the packing arrangement of pedestrian bodies – based on a cross-sectional profile using data from the US National Library of Medicine. Motive torque was taken to arise solely from the pedestrians' orientation toward their preferred destination. The objective was to gain insight into a tragic incident at the Madrid Arena Pavilion in Spain, where five girls were crushed to death. The incident took place at a Halloween Celebration in 2012, in a long, densely crowded hallway used as entrance and exit at the same time. Our simulations reconstruct the mechanics of clogging in the hallway. The hypothetical case of a total evacuation order was also investigated. The result highlights the importance of the pedestrians' density and the effect of counter flow in the onset of avalanches and clogging, and provides qualitative measures of the number of injuries based on a calculation of the contact force network between the pedestrians.

## INTRODUCTION

Pedestrian dynamics modelling has been a subject of mathematical investigation since the 1950s [1; 2], and was originally used by Hankin and Wright to study the movements of pedestrians in the London Subway to allow for more efficient design of stairs and corners. The study of pedestrian panic, however, began later in the 1970s [3] in response to several of history's most horrific crowd stampedes. The most notable of these occurred in Chongqing, China, during World War II. On 6 June 1941 an air raid shelter was evacuated following a Japanese raid, and approximately 4,000 people were killed in a resulting panic-driven stampede. The majority of fatalities were due to asphyxiation [4]. The sites where the Muslim pilgrimage, or Hajj, converge on Mecca have a long history of deaths due to pedestrian panics. When pilgrim luggage spilled over Mecca's narrow Jamarat Bridge in 2006, the crowd was sent into panic; at least 345 people were killed, and 289 were injured [5].

Every year hundreds of people are hurt or lost in crowd disasters. Since Helbing et al. [6] published on intermittent flow and clogging in evacuation processes in 2000, there have been many publications modelling pedestrian behaviour: cellular automata models [7–9], addressing counter flows [7, 8, 10], or force models [11]. Our aim is to identify variables that are helpful for advance warning of crowd disasters: the importance of the shape, density, and counter flow.

In this paper, we present a study of pedestrian dynamics under panic to shed light on a tragic incident in Spain on 31 October 2012, at the Madrid Arena pavilion. The venue was already highly overcrowded when more people entered, and security personnel failed to control the distribution of the crowd, leading to the crushing death of five teenagers when the passageway to the central court was blocked.

A few people attempted to leave the court; but a large number – excited by the music, a desire to be a part of the action, and alcohol – charged in the opposite direction. In the semi-darkness, people were unable to see what lay in front or behind. Security cameras recorded as many as five avalanches of people at the three-meter-wide passageway, the last bottleneck lasting for many minutes. The cameras revealed that one of several fireworks explosions set off more panic, with smoke and sparks it was initially thought that triggered the disastrous final bottleneck. The incident was characterized by shoving, chaos, inability to move in any direction, rising temperatures, claustrophobia, screams, panic attacks, and waning oxygen. Some found themselves lifted off the ground, supported by elbows against their ribs. Those of smaller build, particularly girls, were drawn down under the crush of bodies. Autopsies on the five deceased found suffocation and consequent brain damage as causes of death.

Incidents like these highlight the importance of accurate pedestrian dynamics to inform safe design of hallways, and more specifically in the design of corridors and recommendations on limitations for the movement of people through them. This paper is organised as follows. Section 2 describes the spheropolygon-based model used to perform the simulations; Section 3 describes the Madrid Arena venue and the particular circumstances preceding the tragic incident; Section 4 includes the results of simulations in the hallways were the fatalities happened. We also calculate the cumulative distribution of contact forces in the hallway, and compare it with the distribution in the case of evacuation of the whole venue under panic. Conclusions and discussion of our results are presented in Section 5.

## FORMULATION OF THE PEDESTRIAN DYNAMICS MODEL

The numerical study conducted in this paper is based on a contact force model for pedestrians. The force equations are implemented in a spheropolygon-based model previously used to simulate granular materials [12–14]

**Force model**

The velocity $v_i$ of the pedestrian *i* can be determined according to Newton's second law,

$$\vec{F}_i = m_i \frac{d\vec{v}_i}{dt} \qquad (1)$$

where t is the time, $m_i$ is the mass of the pedestrian, and $\vec{F}_i$ is the force acting on the pedestrian, derived as shown in equation (2):

$$\vec{F}_i = \vec{F}_i^0 + \vec{F}_i^b + \sum_c \vec{F}_{ij}^c \qquad (2)$$

The first two terms are ground reaction forces exerted on the pedestrians by the floor, and $\sum_c \vec{F}_{ij}^c$ is the sum of contact forces produced by other pedestrian and obstacles.

The first term, $\vec{F}_i^0$, denoting the self-driven component of a pedestrian's motion [6]:

$$\vec{F}_i^0 = \frac{m_i}{\tau}(v_0 \vec{e}_i - \vec{v}_i) \qquad (3)$$

Here, $\tau$ is the time required by the pedestrian to reach the maximal velocity and $v_0$ the desired speed. The unit vector $\vec{e}_i$ is the desired direction of motion of the pedestrian.

The third term in equation (2) is the sum of all interactions with other pedestrians, walls, columns or obstacles. Each interaction force of the contact c between particles i and j is calculated as

$$\vec{F}_{ij}^c = \vec{F}_{ij}^{e,n} + \vec{F}_{ij}^{e,t} + \vec{F}_{ij}^{v,n} + \vec{F}_{ij}^{v,t} \qquad (7)$$

where the elastic forces are given by

$$\vec{F}_{ij}^{e,n} = -k_n \delta_{ij}^n \vec{n}_{ij} \qquad (8)$$

$$\vec{F}_{ij}^{e,t} = -k_t \delta_{ij}^t \vec{t}_{ij} \qquad (9)$$

and, $\vec{n}_{ij}$ and $\vec{t}_{ij}$ are the normal and tangential unit vectors. The scalar $\delta_{ij}^n$ is the overlapping length and denotes the vertex-to-edge distance between the two particles. The scalar denoted by $\delta_{ij}^t$ accounts for the tangential elastic displacement given by the frictional force, and satisfies the sliding condition by $\left|F_{ij}^{e,t}\right| \leq \mu F_{ij}^{e,n}$ (10) where $\mu$ is the coefficient of friction. Here, $k_n$ and $k_t$ denote the normal and tangential coefficients of stiffness. The last two terms on the right side of equation (7) account for the coefficient of restitution between pedestrians [14], which are shown in equations (11) and (12).

$$\vec{F}_{ij}^{v,n} = -m_{ij} \gamma_n v_{ij}^n \vec{n}_{ij} \qquad (11)$$

$$\vec{F}_{ij}^{v,t} = -m_{ij} \gamma_t v_{ij}^t \vec{t}_{ij} \qquad (12)$$

where the mass, $m_i = \rho A_i$ (13), the mass of the particle, and $m_{ij} = m_i m_j / (m_i + m_j)$ (14) is the effective mass of the two particles in contact. The density is denoted by $\rho$; and $A_i$ denotes the area of the 2D particle. The normal and tangential coefficients of damping are respectively given by $\gamma_n$ and $\gamma_t$; $v_{ij}^n$ and $v_{ij}^t$ denote the normal and tangential components of the contact velocity.

Additionally, $\vec{F}_i^b$ is the "back force" acting on each pedestrian. This force captures the avoidance behaviour between pedestrians, when pedestrians alter their direction to avoid collision with oncoming pedestrians. It is a ground reaction force, with two components: One is related to the intention of the pedestrians to step to their right to avoid collision, and its direction is the cross product $\vec{e} \times \vec{k}$ where $\vec{k}$ is the unit vector normal to the floor. The second component is a top-up force in the desired direction $\vec{e}$. This force is related to the resistance of the pedestrians to being pushed backwards by the crowd. In situations of panic, when alcohol and crowded hallways may be involved, this back force will be larger than would be observed in normal social conditions.

The back force is important, because it is responsible for stream formation, which is an attribute of crowd dynamics both in social and panic situations. Equation (15) describes the back force:

$$\vec{F}_i^b = \Gamma \Theta(-\vec{e}_i \cdot \vec{v}_i) \frac{m_i v_0}{\tau} \{\vec{e}_i + \vec{e}_i \times \vec{k}\} \tag{15}$$

where the Heaviside step function $\Theta(x)$ returns 1 if x>0, and zero otherwise. This function activates the back force only when the pedestrians are pushed back from their desired direction.

**Shape representation**

(a) 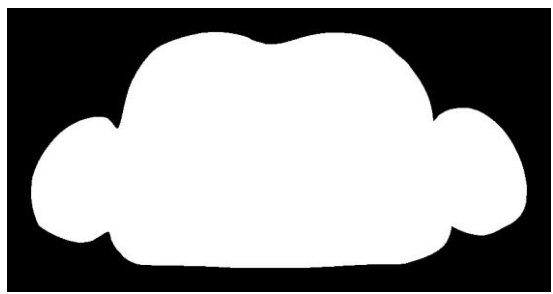

(b) 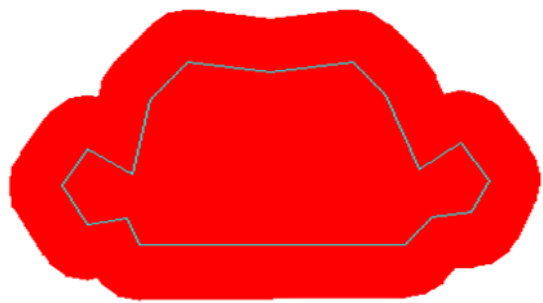

FIGURE 1: (a) Boundary of the cross-section chest derived from US National Library Image of Thorax, and (b) representation of the thorax shape using a spheropolygon with 18 vertices and a sphero-radius of 0.05 m.

Non-spherical shapes have been considered in the pursuit of an ever-more accurate description of pedestrian mechanics [15–18]. In catastrophes involving high densities, the packing arrangement of pedestrians more closely reflects the cross-sectional arrangement of the chest and arms rather than the head. The head has a degree of flexibility, in that it can rotate to avoid collisions, and can fit, from an overhead view, within the area of the thorax and arms. This is consistent with a review conducted from publicly available footage of pedestrian disasters [17–20].

To obtain a more accurate pedestrian shape, a chest cross-section of a human thorax including the arms was obtained from the US National Library of Medicine, which was recoloured in Matlab to isolate the surface. The internal components of the cross section where removed from the recoloured image, and the image was scaled to reflect a typical sternum width of 200 mm. Figure 1 showsthe processed image. This outline is represented using a spheropolygon, which is the figure resulting from sweeping a disk into a polygon, seen in Figure 1b. The advantage of the spheropolygons is that the contact forces among them can be efficiently calculated from their vertex-edge distances, as is explained in [12]

**Torque**

he methods proposed by Korohonen et al. [15–16] and Langston et al. [18] will be used as a foundation for refining the rotational equations of motion.

The ground reaction torque will be considered as arising solely from the pedestrian's desire to face toward their preferred destination, as from [15], applied by the legs at a typical distance of 20 cm. Within the torque which acts on the pedestrian, there is a torque corresponding to the direction of motion, called motive torque, $T_i^m$, and a damping component $T_i^v$.

The force components in the motive torque is applied by the legs, and they allow the pedestrian to rotate in the direction of desired motion. As such, the torque arising from this movement is given by (16), which is a disambiguation of the model constant of the approaches used by Langston et al. and Korhonen et al. [15–16,18]. The λ term in equation (16) is a dimensionless constant that is derived experimentally, $\theta_D = \angle(v_i)$ represents the angle to the walking direction, and θ is the pedestrian's orientation.

$$T_i^m = \frac{\lambda I_i^z}{\tau^2}\left[1-\exp(-\frac{v}{v_0})\right](\theta - \theta_D) \qquad (16)$$

The exponential function is used to account for lower motive torque at lower speeds. This is a reflection of the fact that pedestrians can freely rotate if their velocity is low.

The damping component of the torque reflects the pedestrians desire not to rotate. Using a similar framework to Helbing's self-driven force model [6], in which a desired walking velocity determines the final speed of the pedestrian, a desired angular velocity was created. As the pedestrian prefers not to rotate, the desired angular velocity was set at zero. As a consequence, the pedestrian comes to a comfortable stop after a collision rather than continuing to spin. Equation (17) describes this damping force, where $\tau$ is the relaxation time of rotation, and $I_i^z$ is the moment of inertia:

$$T_i^v = -\frac{I_i^z}{\tau}\omega_i \qquad (17)$$

The net torque acting at time t on pedestrian i, is therefore is described in equation (18):

$$\vec{T}_i = \vec{T}_i^m + \vec{T}_i^v + \sum_{j>i} \vec{\ell}_{ij} \times \vec{F}_{ij}^c \qquad (18)$$

The last term in this equation is the torque produced by the contact forces. The term $\vec{\ell}_{ij}$ is the "branch vector" connecting the center of mass of the pedestrian with the point of application of the contact force.

**Contact parameters**

The parameters of the contact force model were chosen as follows: the thorax responses to forces $k_n = 10^5 \text{N/m}$ are taken from medical data on thorax deformation [21]; $k_6 = 10^4 \text{N/m}$ leads to a bulk Poisson ratio of 0.3; $\mu=0.4$ is close to the coefficient of friction between pieces of cloth fabric. The damping parameter $\gamma_n = 10 \text{s}^{-1}$ was chosen using the formula of the coefficient of restitution [14] between colliding particles, shown in equation (19).

$$\varepsilon = \exp\left(-\frac{\pi \gamma_n / 2}{\sqrt{k_n / m_{ij} - \gamma_n^2 / 4}}\right) \qquad (19)$$

Since not much information is available regarding the coefficient of restitution between pedestrians, a pendulum test was conducted to derive the parameter experimentally. The test was conducted using observations of collisions between pedestrians suspended by ropes by harnesses in a rock climbing venue. By measuring the velocities before and after impact, and neglecting the particle rotation after collision, we obtained coefficients of restitution within the range 0.1–0.5. Details of methods to establish the coefficient of restitution from velocities can be found in the literature on granular media [14].

In the model, a coefficient of restitution of $\varepsilon = 0.3$ was used, which allowed the calculation of the damping parameter $\gamma_n$ from equation (11). The reaction time $\tau=2/3$ s was used, and the terminal velocity was randomly chosen between 0.8 and 1.2 m/s. The surface density of the pedestrians is $\rho=10^3 \text{Kg/m}^2$; mass of the pedestrians were uniformly distributed in the range 40–70 kg. A summary of the key model parameters is provided below in Table 1.

**Numerical solution of the equation of motion**

The in-house object-oriented computer program SPOLY was used to conduct the simulations. SPOLY simulates the pedestrian dynamics based on spheropolygons and a five-order predictor-corrector numerical integration [12]. The use of a neighbour table and Verlet distances allows real-time simulation of long hallways with up to 400 pedestrians. For large-scale simulations – with around ten thousands pedestrians – the algorithm executes around one minute of simulations per hour. Details relating to the SPOLY code are found in [13–14].

TABLE 1 SUMMARY OF MODEL PARAMETERS

| Parameter | Units | Parameter name | Value | Comment |
|---|---|---|---|---|
| $k_n$ | N/m | Normal stiffness | $1 \times 10^5$ | Experimentally calculated using force-displacement relation |
| $k_t$ | N/m | Tangential stiffness | $1 \times 10^4$ | Numerically calculated using elastic analysis of two particles |
| $\mu$ | dimensionless | Coefficient of friction | 0.4 | Obtained from tables of friction between clothes |
| $\epsilon$ | dimensionless | Coefficient of restitution | 0.4 | Experimentally calculated using observations of pendulum test |
| $\lambda$ | dimensionless | Torsion stiffness | 27 | Fitted from comparisson of simulation with video footage. |
| $\Gamma$ | dimensionless | Back force parameter | 1.2 | Fitted from comparisson of simulation with video footage. |

## EMPIRICAL OBSERVATIONS

The insights of this study into crowd panic behaviour are important for the organization of safer mass events as the Halloween party in the Madrid Arena. This is an indoor arena located in Madrid, the capital city of Spain. It is distributed on three floors (access, intermediate, and low). It was planned to house basketball competitions and later featured electronic music artists and a costume contest. The ground floor is the main dance floor, and its dimensions are 55 x 35 m$^2$; see Figure 2.

We recreate the Madrid Arena venue in order to set the scene. The main problem was that the party's organizers were allowed to sell 9,000 tickets – but many more people entered, the judge finding that 19,000 tickets had been issued. Police investigation estimated that during the Halloween party 7,700 of the attendees were on the ground floor. This corresponds to a density of four people per square meter.

(a)
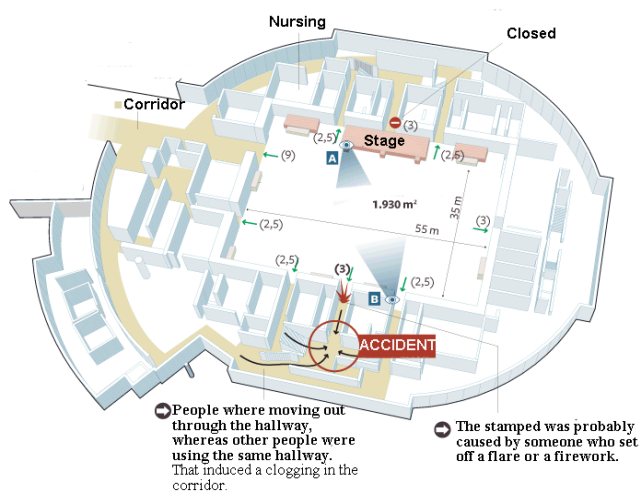

(b)
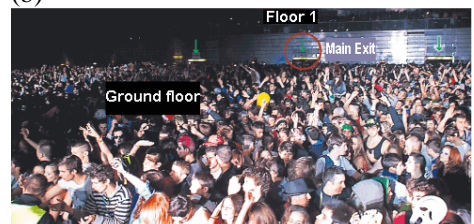

(c)
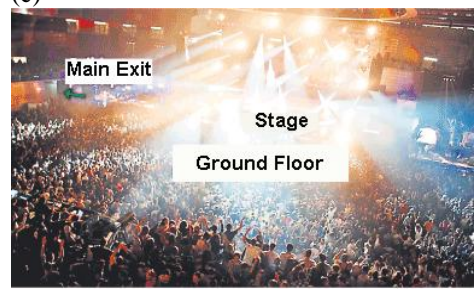

FIGURE 2: (a) Details of the ground floor of the Madrid Arena (ELPAIS, 5/11/2012). (b) The red circle shows the entrance of the hallway were the incident happened. (c) shows the ground floor, the main exit, and the position of the stage.

Police investigation reports a stampede in the main hallway (dimension 12 m x 3 m) that connected the main dance floor to another area (Figure 2a). Initially it was thought that the stampede was caused by someone who set off a flare or a firework nearby, however later it was reported that some firecrackers and some flares were set off in different places – one in the corridor but 20 minutes after the last stampede. When the performance of the invited artist started, the ground floor was packed; however, a crowd of people that were around the venue tried to get in (Figure 3). At this time, some people were trying to get out through the aforementioned corridor. The fact that people were moving out through the hallway, whereas other people were using the same hallway to enter the venue, triggered avalanches along the hallway. Five avalanches were detected before the fatal incident, the last of which ended with people falling down – clogging the hallway, and causing the death of five girls by crushing and asphyxiation.

One witness said the stampede left partygoers piled atop one another as high as their shoulders. Where this fatal incident was happening in the hallway, thousands of partygoers were unaware of the stampede until the police arrived and began to move people out slowly.

Based on the information about crowd density and counter flow taken from video footage, and the available plans of the Madrid Arena venue, we simulated the lane formation, avalanches, and clogging formation in the hallway. We also compared the hallway case with simulations of the hypothetical case of total evacuation of the venue. Evaluation of the contact forces among pedestrians gives us an insight into how the dangerous pressures are building under the crowd conditions in these two cases.

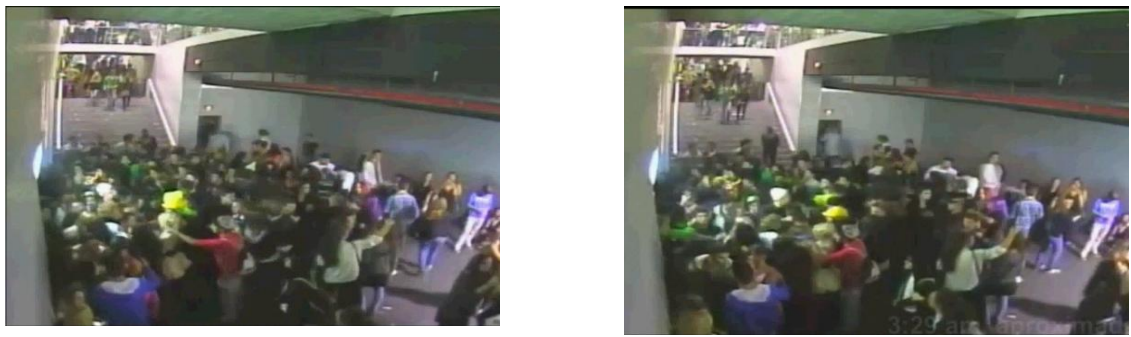

FIGURE 3: Two snapshots from the video footage showing five dangerous avalanches in Madrid Arena preceding the fatal one. These snapshots are of the fifth avalanche, at end of the corridor. The interval between these snapshots is 0.2 seconds, suggesting pedestrian velocities up to 3 m/s. Video footage shows increasing density with time in the hallways. The first avalanches happened at 2:18 am, 2:19 am, 2:41 am, 2:59 am, and 3:29 am. In the fatal incident, registered at 3:33 am, people fell down, clogging the hallway.

## RESULTS

**Simulation of the hallways**

In the first series of simulations pedestrian flow in the 3 m x 12 m hallway was investigated, with replication of Madrid Arena's corridor. A random distribution of pedestrians was placed in the simulation area. Following this, a percentage of the pedestrians were assigned a desired direction towards the exit, and the others were assigned a direction towards the venue to represent the counter flow. With the aim to replicate observations from the Madrid Arena venue, these simulations were repeated using different densities. Periodic boundary conditions were used in the edges of the corridor.

Depending on the density of pedestrians and the counter flow percentage, three different regimes are observed in the simulations, see Figure 4: i) lane formation, which occurs at low density and is characterized by well-defined lines of pedestrian flow at the stationary stage; ii) avalanches, which are observed at intermediate densities and counter flow values, characterized by temporal clogging of the pedestrian flow, followed by a sudden release of potential energy that leads to a short-time accelerating flow; and iii) clogging, characterized by a frozen stage where the pedestrians are clogged indefinitely in the corridor.

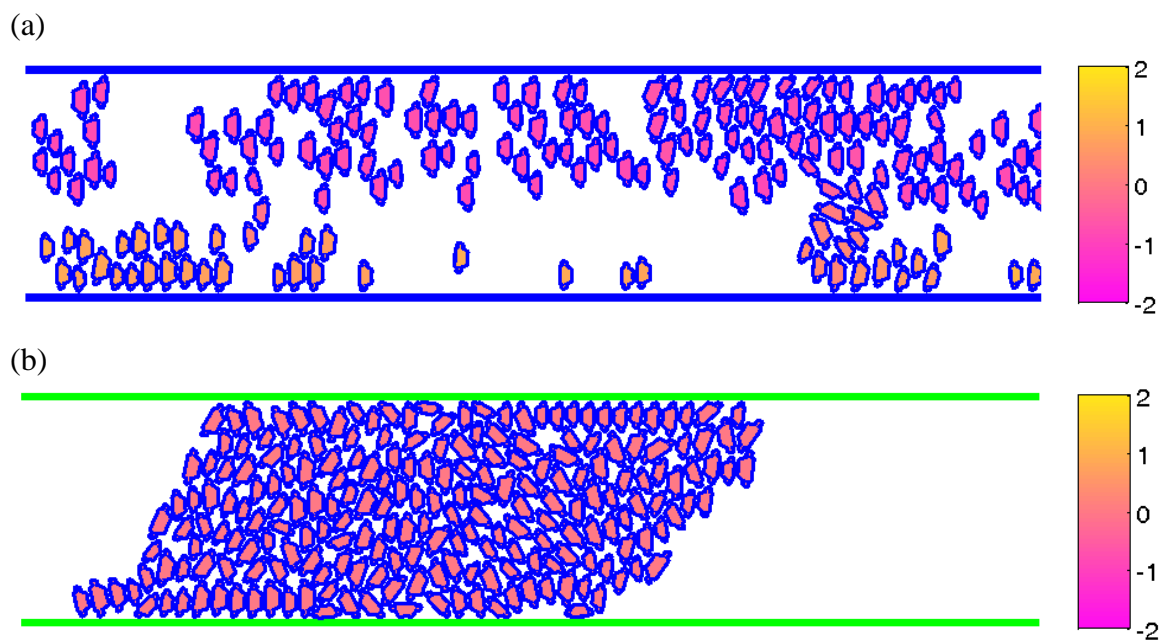

Figure 4: Two different regimes observed inteh simulation of the hallway.. A snapshot of th lane formation is shown in (a) and a snapshot with the clogging is shown in (b).

Figure 5 shows the phase diagrams of these three regimes. The avalanche regime is sandwiched between the clogging and the land-formation regime. Below 4.4 pedestrians/m$^2$ and below a counter flow of 20%, the model predicts a formation of lanes in the stationary stage. The formations of avalanches without clogging appear up to a pedestrian density of 5.8 pedestrians/m$^2$, and a counter flow up to 34%. These avalanches are critical in pedestrian safety as they can produce people falling down and clogging entire hallways. The most dangerous situations are observed above 5.8 pedestrians/m$^2$, and counter flow between 34% and 66%. In these cases the model predicts complete clogging, which leads to long-lasting pressure among the pedestrians and, if the pressure are too high, possible injury by asphyxiation.

It is interesting to note that the phase diagram depends on the pedestrian shape used in the model. Thoracic shapes leads to more conservative results than circular shapes; in the later it suggests clogging above 6.3 pedestrians/m$^2$. The avalanches in the thorax model are less pronounced than in the model of circular particles. This is probably due to the shoulder rotation of the thoracic-shape pedestrians, which reduces the accumulation of stress that produces avalanches and eddy-like motion in the boundaries between lanes. We should note that the resistance of the pedestrians to rotation or to being pushed back depends on several psychological conditions and cultural background. These are accounted by the parameters $\Gamma$ and $\lambda$ of Eqs. 15 and 16 in our model, which has been fitted from comparison of simulations with video footage observations. Our simulations show that increasing the parameter $\Gamma$ of the back force reduce the zone of avalanches in the phase diagrams, and increasing the torsion stiffness parameter $\lambda$ increases the zone of clogging.

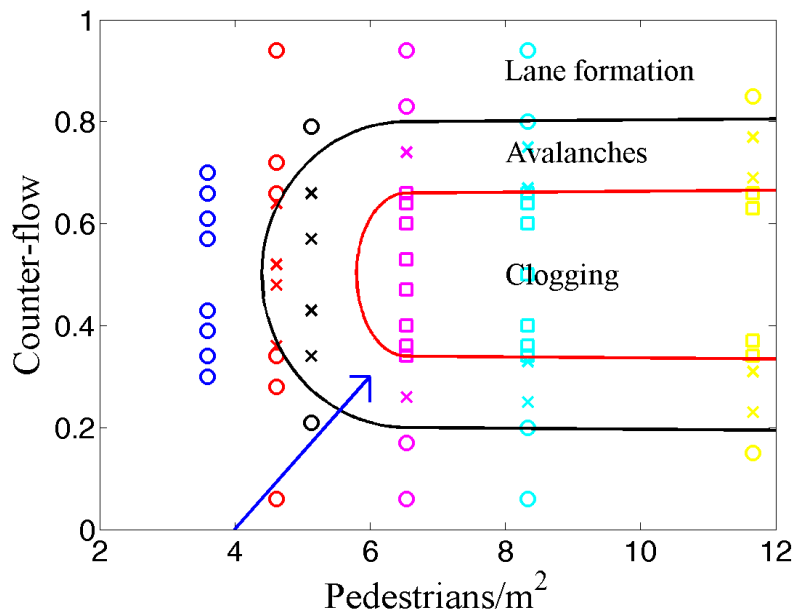

FIGURE 5: Phase diagram of the simulations of counter-flow in the corridor with circles. The regimes observed are: lane formation (disks), avalanches (crosses) and clogging (squares). The arrow shows the time evolution of these two parameters in the hallways in the Madrid Arena. It started form a dense crowd of people leaving the party. Then a counter-flow of people joining the event increased the density, and finally it produced the avalanches and eventually the fatal clogging.

## Cumulative size distribution of forces and fraction of injuries

### Risk of asphyxia by crushing

There is a lack of specific data concerning this type of force in these situations that cause asphyxia by crushing. Some data are found from major crowd catastrophes [22] such as those in Bolton in 1946 [23], Glasgow in 1971 [24], or Sheffield in 1989 [25]. Smith and Lim [26] carried out an experimental investigation of the loads that people can stand when being pushed against a barrier.

A lower-bound comfort level level that could be withstood for 30 seconds or so without major distress was determined. In previous studies Anon (undated) at the University of Surrey [27] pulled the subject against a barrier by means of cords attached to a pack frame on the subject's back. Part of these tests attempt to measure a subjective load level of discomfort estimated by the subject to become intolerable after a few seconds. The mean load determined at chest level from 17 subjects was 418 N with a minimum of 116 N and a maximum of 774 N. Hopkins et al. [28] also measured pressures in a public event. The data were collected during an 85-minute performance by the English band Ride, at a heavy rock concert in Britain. These crowd pressures were measured from the load measuring transducers attached on eight panels arranged in front of the stage, each of which was of 1.5 m length.

It is difficult to assess the load and the time needed to produce asphyxia in the five girls at the Madrid Arena; however, taking into account the visual observation of the incident a reasonable estimation of the load might be 400 N, considering the number of stacked pedestrians. These data will be used in our simulations to provide estimates of the number of people at risk.

**Cumulative size distribution**

Two different cumulative size distribution scenarios were considered: the clogged hallway and the entrance in the case of panicking evacuations.

Video footage from the Madrid Arena disaster showed that after five human avalanches in the hallway some pedestrians fell down, which subsequently created a blockage near the exit of the corridor. To simulate this blockage some additional rules were included in the dynamics regime: if pedestrians accelerate above $5m/s^2$ they were assumed to fall down. A fallen pedestrian was assumed to behave like an obstacle; that is, they cannot either rotate or translate. A snapshot of the simulation of this clogging is shown in Figure 6 below. The fallen pedestrians create bottleneck conditions, which lead to a stable clog. At this stage, most of the pedestrians enter into physical contact, creating a contact network. We represent this contact network by drawing a line from the center of mass of the pedestrian to the contact point, and encoding the magnitude of the force by the thickness of the line.

The contact network between the particles, shown in Figure 6 as an example, demonstrates that the forces on particles are strongly heterogeneous, as their distribution is not uniform but organized in the form of filamentary structures that granulologists call "force chains". When applied to pedestrian motion, these force chains can determine which pedestrians suffer the largest forces and are most likely to choke. At any given point, a pedestrian may bear enormous forces, while an adjacent pedestrian may experience almost no force. In Figure 6 we also see the percentage of pedestrians who are save, uncomfortable, and in danger.

A dramatic distribution of contact forces is observed in the hypothetical case of an evacuation in a large room filled with pedestrians. A simulation was conducted using an evacuation of 6,800 pedestrians in a room with the same dimensions of the Madrid Arena. By calculating the contact network we were able to investigate theoretically the severity of the contact forces on each pedestrian and demonstrate the number of pedestrians at severe risk. Due to the stress supported by the walls, the forces built in a hallway such as the corridor of the Madrid Arena are not large enough for many pedestrians to accumulate (in the order of ~10), but due to the large deviations in the force distribution we would expect few fatalities. Besides, due to the large duration of this clogging in the hallways (several minutes) the strongest forces were enough to produce asphyxiation in a few pedestrians, accounting for the five fatalities on Halloween in the Madrid Arena.

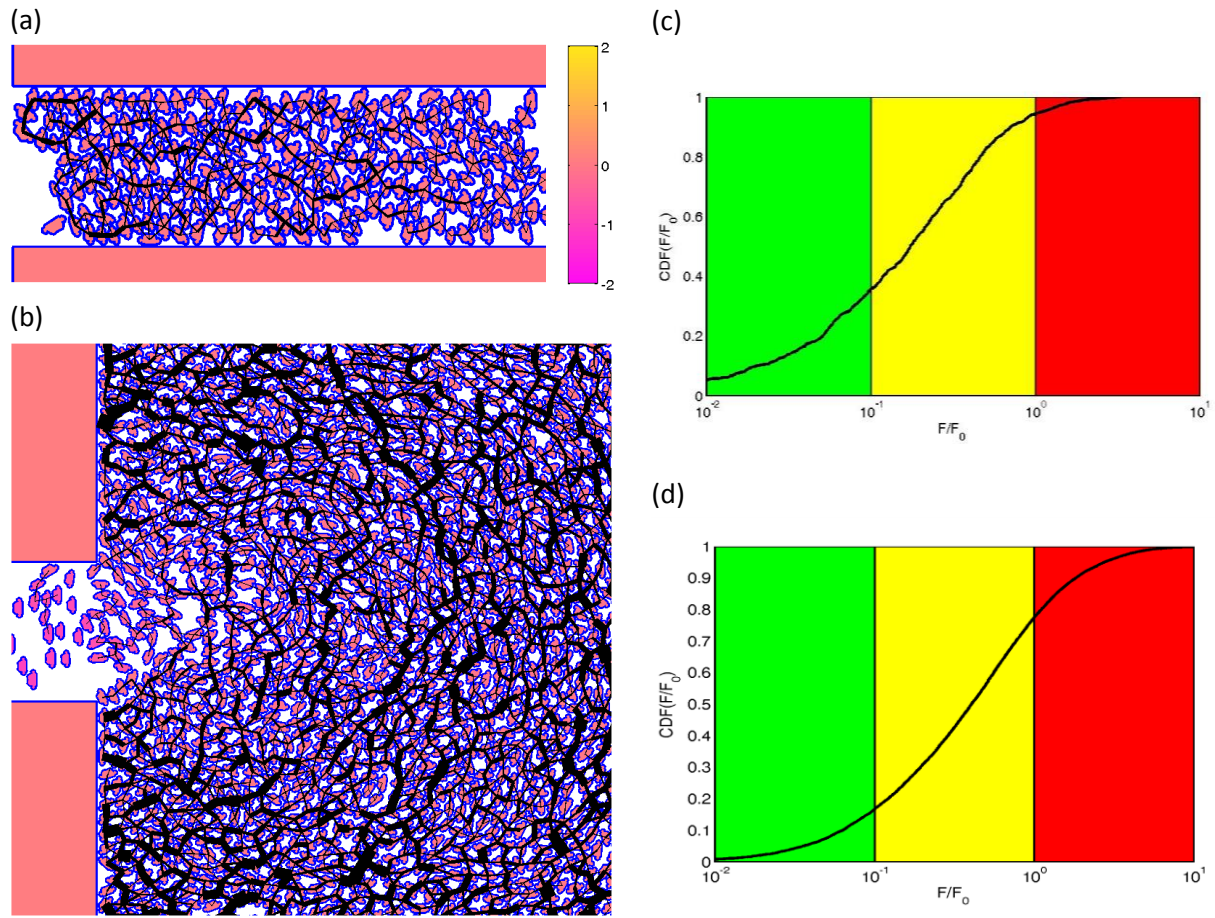

FIGURE 6: Contact force network in the Madrid Arena. (a) shows the contact forces during the blockage incident in the hallway of the Madrid Arena; the thickness of the line encodes the magnitude of the contact force. (c) shows the cumulative distribution of forces (CDF) scaled by $F_0=400N$. (b) shows the contact force during the panicking evacuation of the venue (hypothetical). The CDF of this case is shown in (d). The colours in (c–d) represent the safe zone (green), uncomfortable zone (yellow), and the danger zone (red). The percentage of fatalities in the total evacuation case is five times larger than in the blocked corridor.

## CONCLUSIONS

We presented a pedestrian interaction model based on contact forces, ground reaction forces and torsion, and representation of thorax shapes using spheropolygons. The model is suitable for calculations of flow rates and contact force network under crowd conditions, where pedestrians interact more by chest and arm arrangement than by avoiding physical contact. For less dense scenarios where the private space of each pedestrian is relevant, the model may require the introduction of long-range social force as proposed by Helbing [6].

The flowability of the pedestrians in long hallways was modelled as a function of the density of pedestrian, and the percentage of pedestrian in counter flow. We obtain three regimes: lane formation, avalanched flow and clogging. These regimes allowed us to reconstruct the tragedy of the Madrid Arena. From our simulations we found that the crowd in the hallway turned from a lane formation regime to the avalanches regime due to increase of density and the counter-flow. These transitions suggest that the Madrid Arena incident may have been produced by an ever-increasing crowded counter-flow condition rather than by panic derived by firework, as suggested by early official reports–Indeed after the videos the police report claimed that the firework was 20 min after the last and fatal avalanche.

A particularly difficult aspect of the modelling was simulating the people falling in the floor. We assumed that these people act as obstacles for the other pedestrians. This is a quite crude simplification, as simplifies a three dimensional problem in a two dimensional model. Yet our two dimensional model allowed us to calculate the force experience by the pedestrians, and to estimate a small, but important, fraction of pedestrians who suffered from long-lasting pressure that derived into asphyxiation.

We noted from the reports on the incident that the authorities allowed the party to continue in spite of the tragedy. Simulations showed that if further panic was caused by hastened evacuation, it might have produced a bottleneck situation in the entrance of the corridor. Under these conditions, our model predicts that the number of causalities would be five times larger.

The force distribution found in our simulations was quite heterogeneous, suggesting that it is organizing in force chains that carry more of the load produced by the crowd. These force chains carry large forces and persist for several minutes, potentially leading to crushing and asphyxiation. It remains unclear whether an individual pedestrian could avoid been trapped by these force chains by moving their arms and rotating their chest (Actually there is an emergency procedure in these cases to help against asphyxia: to cross the arms, left hand in right shoulder, right hand in left shoulder). Under dense crowd conditions, the pedestrian would have little space and would therefore find it difficult to re-arrange themselves. Even if a single force chain is temporarily dissipated by a pedestrian escaping, the large pressure from other panicking pedestrians would lead to the force chain being re-established.

Two main parameters were identified in the contact force model: the coefficient of friction, which is the main controller of the flow rate in dense granular flow, followed in importance by the coefficient of restitution. Given the importance of these parameters it is recommended that further empirical work be conducted to provide greater model accuracy. Ethical issues can be avoided by using sand inside stitched clothes or crash-test dummies. The test dummies would have the convenience of already having inbuilt sensors to measure pressure-deformation relations.

Psychological parameters that need to be investigated in further studies include the $\lambda$ factor related with the stiffness of the shoulder rotation, and the $\Gamma$ factor related to strength of the back force used by pedestrians to avoid being push back from their desired direction. We presume that these parameters depends on cultural aspects and special circumstances of the crowd, and they would need to be calibrated by observations of crowd dynamics under similar conditions before making predictions. Further studies should focus in providing recommendations of design for hallways and solve organizational issues for this type of events looking for crowd safety.

## ACKNOWLEDGMENTS

C. L. thanks Asociacion de Amigos la Universidad de Navarra for a scholarship. F.A.M. acknowledges discussion with Iker Zuriguel, and Daniel Parisi, and technical support from Candace Lu, Shumiao Chen, Fiona Tang Huey Ming, and Guien Miao.